\documentclass[a4paper,10pt]{article}
\usepackage{graphicx}
\usepackage{graphics}
\title{\huge\bf Estiamte of the two-photon exchange effect 
on deuteron electromagnetic form factors}
\author{\bf Yu Bing Dong \\
Institute of High Energy Physics, Chinese Academy of Sciences,\\
Beijing 100049, P. R. China\\
and\\
Theoretical Physics Center for Science Facilities (TPCSF), 
CAS, P. R. China}
\begin{document}
\maketitle
\begin{abstract}
The corrections of two-photon exchange on deuteron electromagnetic form 
factors are estimated based on an effective Lagrangian approach. 
Numerical results for the form factors $G_{C,M,Q}$ of the deuteron with the 
corrections are compared to its empirical ones. Moreover, the two new 
form factors, 
due to the two-photon exchange, are analyzed. Possible way to test the 
two-photon exchange corrections to the deuteron form factors is discussed. 

\end{abstract}

\newpage

\section {Introduction}

{\hskip 0.4cm}We know that the electromagnetic (EM) form factors of the proton 
and deuteron are usually extracted from the measurements of the 
differential cross sections  of $ep$ and $eD$ elastic scatterings  
and from the Rosenbluth separation method \cite {Rosen}, which is based on 
one-photon-exchange (OPE) approximation. For a long time, the 
extracted $Q^2$-dependences of the nucleon EM form factors are believed to 
behave like a simple dipole form. For the proton electric and magnetic form 
factors, $G_{E,M}^p$, one conventionally assumes   
\begin{eqnarray}
G^p_E(Q^2)=G^p_M(Q^2)/\mu_p\simeq 1/(1+Q^2(GeV^2)/0.71)^2,
\end{eqnarray}   
where $\mu_p=2.79$ is the proton magneton. Recently, the new experiments of 
the polarized $ep$ elastic scattering were precisely carried out 
at Jefferson Laboratory \cite {Jones}. The polarization transfer scattering 
experiments of $\vec{e}+p\rightarrow e+\vec{p}$ show that the ratio 
$R^p=\mu_pG_E^p(Q^2)/G_M^p(Q^2)$ behaves like $R^p(Q^2)\sim 1-0.158Q^2$. 
It means that $R^p$ is no longer a simple constant as implied in eq. (1). 
It monotonously decreases with the increasing of $Q^2$. \\

One way to resolve this discrepancy, at least partially, is to take the 
effect of the two-photon-exchange (TPE) into account \cite {Gui, Blu, Chen, 
GT, Arr, Gun}. Usually, it is believed that TPE is strongly suppressed 
by EM coupling constant $\alpha_{EM}$ ($\sim 1/137$).  However, it was 
argued \cite {Gun} 
that due to a very steep decreasing of the nucleon EM form factors, the 
TPE process, where the $Q^2$ is equally shared by the two exchanging photons, 
may be compatible to the OPE one. Some calculations of the TPE corrections  
to the $ep$ elastic scattering have been done recently \cite{Gui, Blu,  
Chen, GT, Arr, Blu0}, where only the nucleon state is considered as 
an intermediate state. The calculations were extended further with other 
nucleon resonances, like $\Delta$, $P_{11}$ and $D_{13}$ states, being 
considered   as the intermediate states  \cite {Kond}. There were also 
several other 
works about the TPE effect on the proton charge radius and on the 
parity-violating \cite {Car, Blu1} in the $ep$ scattering. The effect on 
the EM form factors of the nucleon in the time-like region was estimated 
in Refs. \cite {Dong3, GT4}. According to the analyses for the TPE effect on 
the nucleon EM form factors in the literature, it is known that the TPE 
corrections not only modify the conventional nucleon electric and magnetic 
form factors, but also provide a new form factor, $Y_{2\gamma}$, to the 
nucleon.\\
 
The TPE corrections to the deuteron (spin 1 
particle) EM form factors and to the $e^{+}+e^-\rightarrow D+\bar{D}$ 
process have been also discussed in Refs. \cite {Dong, Ga5, Ga6} 
qualitatively. In analogy to the TPE effect on the proton EM form factors, 
TPE not only modifies the conventional three EM form factors of the 
deuteron, but also provides new form factors with new structures. The 
general discussion of the structures of the three new form factors can be seen 
\cite {Dong, Ga5}. We know that the deuteron is usually 
regarded as a weekly bound system of a proton and a neutron (see Fig. 1). 
Many calculations for the EM form factors of the deuteron, with the OPE 
approximation, have been performed in different approaches in the literature 
(see for example, Refs. \cite {mathiot,gari,karmanov,kaplan}). Recent 
calculations based on an effective Lagrangian approach \cite {Ivanov,Dong0} 
have shown that this approach can reasonably explain the deuteron 
EM form factors with phenomenological including two-body operators. \\

\begin{figure}[t]
\centering \includegraphics[width=8cm, height=4cm]{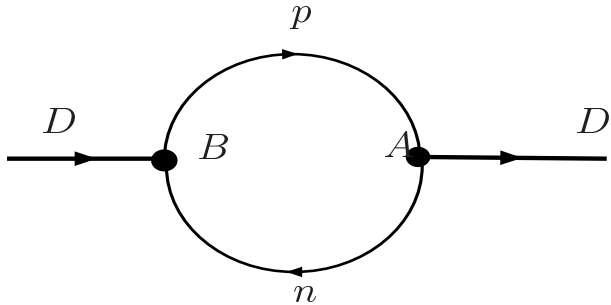}
\caption{\footnotesize Deuteron mass operator.}
\end{figure}

To study the TPE effect on the deuteron system in our effective Lagrangian 
approach, we note that the deuteron EM form factors receive the TPE 
corrections from three different sources.  The first one 
(see Fig. 2) is that the two photons directly couple to the contact points 
(the contact point A (or B) of Fig. 1 is the one connects the deuteron to its 
composites). The second is that one of the two photons directly couples to 
one of the nucleons and another to one of the contact points (see Fig. 3).
The last one is that the two photons respectively couple to the two nucleons 
(see Fig. 4). It has been proved that gauge invariance preserves in our 
effective Lagrangian approach only when the three kinds of the 
two-photon exchange diagrams are considered simultaneously 
\cite {faesslernew}. \\
  
In our previous work \cite {Dong4}, only a part of the third type 
of the TPE corrections to the EM form factors of deuteron is considered, 
where the TPE corrections to the EM form factors $G_{E,M}$ and to 
$Y_{2\gamma}$ of the proton and neutron are directly employed to study the 
deuteron properties following the formalism of Ref. \cite {Blu0}. In the 
approach, only one new form factor appears. The TPE effect considered in 
\cite {Dong4} is represented  by Figs. 4(a), 4(b) and their cross-box 
diagrams. In this paper, to extend the work of \cite{Dong4} further, 
we'll simultaneously study the three sources of the TPE  effect in Figs. 2-4. 
It should be stressed that although the contribution of the coupling of a 
photon to the contact point is expected to be smaller than the one of  
direct couplings of the photon to the nucleons, this type of couplings is 
needed in order to guarantee gauge invariance. This paper is organized as 
follows. In section 2 the above mentioned two-photon-exchange effect in 
the $eD$ elastic scattering is briefly discussed. Numerical results  and 
conclusions are given in section 3. \\

\begin{figure}[t]
\centering \includegraphics[width=15cm,height=5cm]{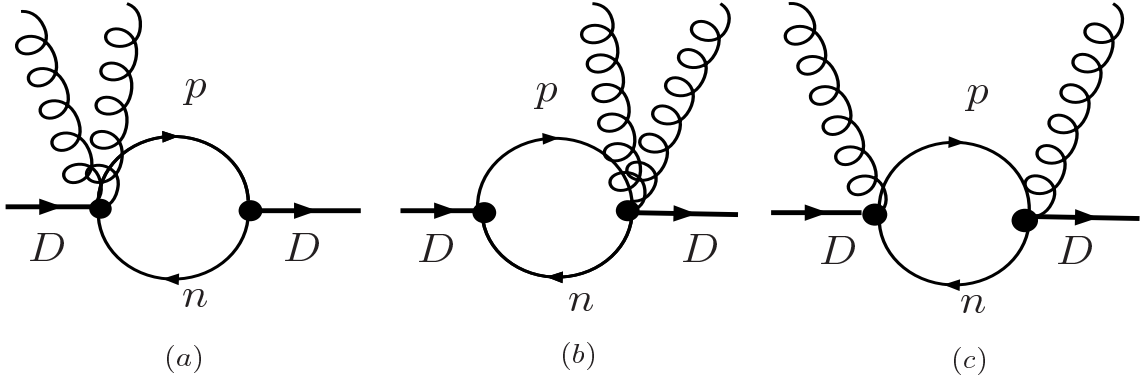}
\caption{\footnotesize Diagrams for the first type of the two-photon 
exchange effect. The cross-box diagrams are implied.}
\end{figure}

\section {Two-Photon-Exchange in the $eD$ elastic scattering}

{\hskip 0.4cm}According to the OPE approximation, the electromagnetic form 
factors of the deuteron are defined by the matrix element of the 
electromagnetic current $J_\mu(x)$ 
\begin{eqnarray}
&&<p'_{D},~\lambda'\mid J_\mu(0)\mid p_D,~\lambda> =-e_D\bigg\{\Big
[G_1(Q^2) \xi'^*(\lambda')\cdot \xi(\lambda)\\ \nonumber 
&&-G_3(Q^2)\frac{(\xi'^*(\lambda')\cdot q)(\xi(\lambda)\cdot
q)}{2M^2_D}\Big ]\cdot P_\mu 
+G_2(Q^2)\Big [\xi_{\mu}(\lambda)(\xi'^*(\lambda ')\cdot q)-
       \xi'^*_{\mu}(\lambda ')(\xi(\lambda)\cdot q)\Big ]\bigg \},
       \label{current}
\end{eqnarray}
where $p'_{D},\xi',\lambda'$ (or $p_D,\xi,\lambda$) denote the
momentum, helicity, and polarization vector of the final (or initial)
deuteron, respectively. In eq. (2) $q=p'_{D}-p_D$ is the photon momentum,
$P=p_D+p'_{D}$, $Q^2=-q^2$ is the four-momentum transfer squared, $M_D$
is the deuteron mass, and $e_D$ is the charge of the deuteron. In the
one-photon exchange approximation or Born approximation, the
unpolarized differential cross section of the $eD$ elastic scattering,
$e(k_1,s_1)+D(p_D,\xi)\rightarrow e(k'_1,s_3)+D(p'_{D},\xi')$,
in the laboratory frame is~\cite{Jankus56}
\begin{eqnarray}
\frac{d\sigma}{d\Omega}&=&\frac{d\sigma}{d\Omega}\bigg
|_{Mott}I_0(OPE), \label{diffCrx}\nonumber \\
I_0(OPE)&=&A(Q^2)+B(Q^2)tan^2\frac{\theta}{2}, 
\end{eqnarray}
where $\theta$ is the scattering angle of the electron, 
$(d\sigma/d\Omega)_{Mott}$ is the Mott cross section for a structure-less 
particle with recoil effect, and the two structure functions are
\begin{eqnarray}
A(Q^2)&=&G_C^2(Q^2)+\frac23\tau_D
G_M^2(Q^2)+\frac89\tau^2_DG_Q^2(Q^2),\nonumber \\
B(Q^2)&=&\frac43\tau_D(1+\tau_D)G^2_M(Q^2). \label{AB}
\end{eqnarray}
In eq. (4) $\tau_D=Q^2/4M^2_D$, and $G_M$, $G_C$ and $G_Q$ are the deuteron 
magnetic, charge and quadrupole form factors, respectively. They can be 
expressed, in terms of $G_1$, $G_2$ and $G_3$, as 
\begin{eqnarray}
G_M=G_2, ~~~~G_Q=G_1-G_2+(1+\tau_D)G_3,~~~ G_C=G_1+\frac23\tau_D G_Q.
 \label{GMCQ_G123}
\end{eqnarray}
The normalizations of the three form factors are  
$G_C(0)=1$, $G_M(0)=1.714$, and $G_Q(0)=M^2_DQ_D=25.83$. Note that in eqs. (3) 
and (4), there are two unpolarized structure functions $A$ 
and $B$, and three independent form factors $G_C$, $G_Q$ and $G_M$ for the 
deuteron. To determine the three form factors completely, one needs, at 
least, one polarization observable. The optimal choice is the polarization 
$T_{20}$ (or $P_{zz}$) \cite {Gar1}.\\

\begin{figure}
\centering \includegraphics[width=10cm,height=8cm]{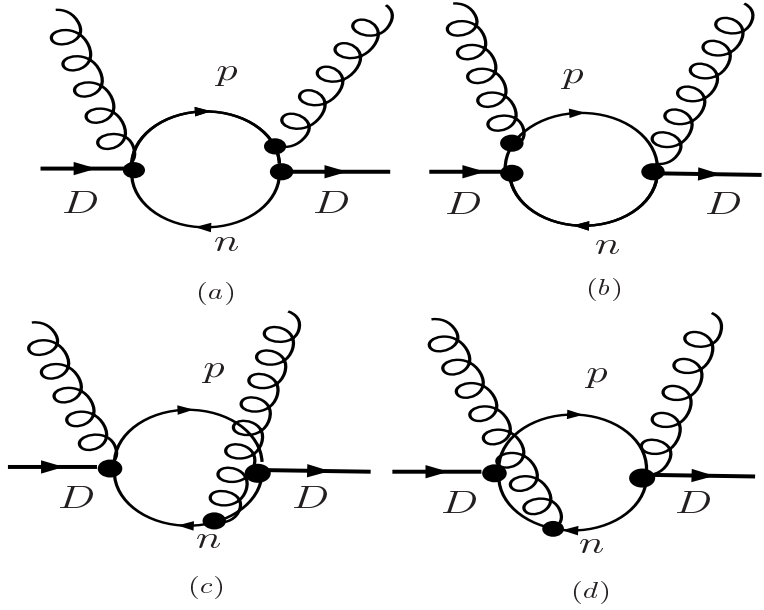}
\caption{\footnotesize Diagrams for the second type of the two-photon 
exchange effect. The cross-box diagrams are implied.}
\end{figure}

Considering both OPE (${\cal C}=-1$) and TPE (${\cal C}=+1$), and  taking 
Lorentz, party, and charge-conjugation invariance into account, one obtains 
the most general form of the $eD$ elastic scattering \cite{Dong, Tarrach75},
\begin{eqnarray}
{\cal M}^{eD}=\frac{e^2}{Q^2}\bar{u}(k'_1,s_3)
\gamma_{\mu}u(k_1,s_1)\sum_{i=1}^6G_i'M_i^{\mu},
\label{MeD}
\end{eqnarray}
where
\begin{eqnarray}
M_1^{\mu}&=&(\xi'^*\cdot\xi)P^{\mu},~~~\nonumber \\
M_2^{\mu}&=&\Big [\xi^{\mu}(\xi'^*\cdot q)
-(\xi\cdot q)\xi'^{*\mu}\Big ],\nonumber \\
M_3^{\mu}&=&-\frac{1}{2M_D^2}(\xi\cdot q)(\xi'^*\cdot q)P^{\mu},~~~
\end{eqnarray}
and 
\begin{eqnarray}
M_4^{\mu}&=&\frac{1}{2M_D^2}(\xi\cdot K)(\xi'^*\cdot K)P^{\mu},\nonumber \\
M_5^{\mu}&=&\Big [\xi^{\mu} (\xi'^*\cdot K)
+(\xi \cdot K)\xi'^{*\mu}\Big ],\nonumber \\
M_6^{\mu}&=&\frac{1}{2M_D^2}\Big [(\xi\cdot q)(\xi'^*\cdot K)
-(\xi\cdot K)(\xi'^{*}\cdot q)\Big ]P^{\mu},
\end{eqnarray}
where $K=k_1+k'_1$. General speaking, the form factors $G'_i$, 
with  $i=1,6$, are complex functions of $s=(p_D+k_1)^2$ and
$Q^2=-(k_1-k'_1)^2$. They can be expressed as
\begin{eqnarray}
G_i'(s,Q^2)=G_i(Q^2)+G_i^{(2)}(s,Q^2), \label{Gtilde}
\end{eqnarray}
where $G_i$ corresponds to the contributions arising from the
one-photon exchange and $G_i^{(2)}$  stands for the rest which
would come mostly from TPE. In the OPE approximation,
$G_4'=G_5'=G_6'=0.$ It is easy to see that $G_i$ $(i=1,2,3)$ are of
order of $(\alpha_{EM})^0$ and $G_i^{(2)}$ ($i=1,...6$) are
of order  $\alpha_{EM}$. \\

\begin{figure}[t]
\centering \includegraphics[width=10cm,height=9.0cm]{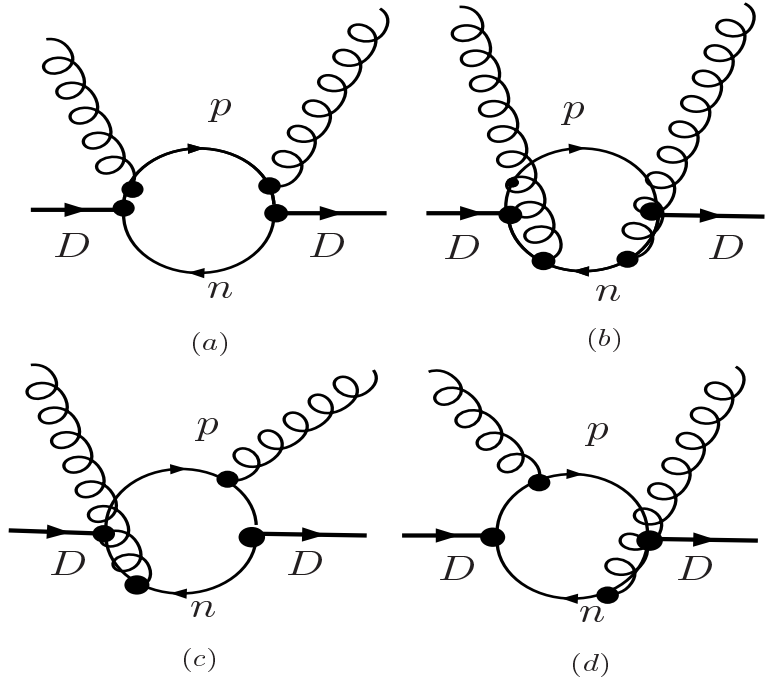}
\caption{\footnotesize Diagrams for the third type of the two-photon 
exchange effect. The cross-box diagrams are implied.}
\end{figure}

To consider that a deuteron is a weakly bound state of a proton and a 
neutron, we take the following effective interaction between the deuteron and 
its composites ($pn$) 
\cite{Dong0}
\begin{eqnarray}
{\cal L}_D = g_{D}D^{\mu+}(x)\int dy 
\Phi_D(y^2)\bar{p}(x+\frac12y)C\gamma_{\mu}
n(x-\frac12y)+ {\rm H.c.},
\label{ls}
\end{eqnarray}
where $C$ is the charge conjugate matrix, $D^{\mu}$, $p$ and $n$ are the 
fields of the deuteron, proton and neutron. The correlation function $\Phi_D$ 
in eq. (10) characterizes the finite size of the deuteron as a $pn$ bound 
state and depends on the relative Jacobi coordinate $y$, in addition, $x$ 
being the center-of-mass (CM) coordinate. The Fourier transformation of the 
correlation function reads 
\begin{eqnarray}
\Phi_D(y^2) \, = \, \int\!\frac{d^4p}{(2\pi)^4}  \,
e^{-ip y} \, \widetilde\Phi_X(-p^2) \,. 
\end{eqnarray}
A basic requirement for the choice of an explicit form of the correlation
function is that it vanishes sufficiently fast in the ultraviolet region
of Euclidean space to render the Feynman diagrams ultraviolet finite.
Here, we adopt a Gaussian form 
$\tilde\Phi_D(p_E^2) \doteq \exp( - p_E^2/\Lambda_D^2)$
for the vertex function, where $p_{E}$ is the Euclidean Jacobi momentum of
 the deuteron, and $\Lambda_D$ is a size parameter. It characterizes the 
distribution of the constituents inside the deuteron.\\

We know that the low energy theorem \cite {LET0} provides a model independent 
test for the reliability of different approaches \cite {LET1}. For the 
photon deuteron (spin-1) Compton scattering, the low energy theorem has been 
discussed extensively in the past \cite{LET2}. A complete treatment 
on this issue has been given by Ref. \cite {LET3}.  To the first order 
of the photon energy $\omega$, the low energy theorem tells that the forward 
Compton scattering amplitude off the deuteron target is \cite{LET1}
\begin{eqnarray}   
4\pi{\cal T}=-\frac{e^2}{M_D}\vec{\epsilon}~'\cdot\vec{\epsilon}
-i\frac{e^2}{4M_D^2}\omega(\mu_D-2)^2\vec{S}\cdot
(\vec{\epsilon}~'\times\vec{\epsilon})+{\cal O}(\omega^2),
\end{eqnarray}
with $\vec{\epsilon}$ (or $\vec{\epsilon}~'$), $\vec{S}$ and $\mu_D$ being 
the initial (or final) photon polarization, the deuteron spin and its 
magnetic moment in unit of $e/2M_D$, respectively. In eq. (12) the first 
and second terms are the Thomson and the spin-flip ones. The latter is 
proportional to the deuteron anomalous magnetic moment squared 
$\kappa_D^2=(\mu_D-2)^2$ and associates to the well-known 
Drell-Hearn-Gerasimov sum rule of the deuteron (see \cite {LET1,LET4} for 
example). In our effective Lagrangian approach \cite {Dong0}, the effective 
current of photon-deuteron has the correct structures like eq. (2) and the 
numerical calculation shows that the obtained magnetic moment of the deuteron 
$\mu_D$ is around $1.7$ (in unit of $e/2M_D$), which reasonably agrees with 
the experiment data. Consequentially, it is expected that the forward Compton 
scattering amplitude based on our effective approach is consistent with the 
low energy theorem. It should be mentioned that the above correlation 
function of eq. (11), in the non-relativistic approximation, stands for the 
wave function with only S-wave of the deuteron, which does not contain any 
D-wave component. To reasonably explain the data for the deuteron quadrupole 
moment, we have to phenomenological include two-body operators \cite{Dong0}. 
A detailed comparison of the photon-deuteron Compton scattering 
amplitudes in the low photon energy region of our approach and of the low 
energy theorem will be explicitly given in a our separate paper. \\

In our approach, the coupling $g_D$ of 
$<p^D,\lambda \mid pn> =g_D\xi^{'*}(\lambda)$ 
is determined by the compositeness condition~\cite{faesslernew, Weinberg,
Efimov,Anikin,Faessler}. It implies that the renormalization constant of 
the deuteron wave function is set equal to zero:
\begin{eqnarray}
\label{ZX}
Z_D = 1 - \Sigma^\prime_D(M_D^2) = 0.
\end{eqnarray}
Here, 
\begin{eqnarray}
\Sigma^\prime_D(M_{D}^2)=
g_{_{D}}^2 \frac{d\Sigma_D}{dp_D^2}\Huge{\mid}_{_{p_D^2=M_D^2}}
\end{eqnarray} 
is the derivative of the transverse part of the mass operator
$\Sigma^{\alpha\beta}_D$, which  conventionally splits into the transverse
$\Sigma_D$ and longitudinal $\Sigma^L_D$  parts as:
\begin{eqnarray}
\Sigma^{\alpha\beta}_{D} = g^{\alpha\beta}_\perp \Sigma_D(p^2_D) 
+ \frac{p_D^\alpha p_D^\beta}{p_D^2} \Sigma^L_D(p^2_D) \,,
\end{eqnarray}
where 
$g^{\alpha\beta}_\perp = g^{\alpha\beta} - p^\alpha p^\beta/p^2\,, and 
\hspace*{.2cm} g^{\alpha\beta}_\perp p_\alpha = 0\,.$ 
The mass operator of the deuteron in our approach is described by 
Fig. 1.  If the size parameter $\Lambda_D$ 
is fixed, the coupling $g_D$ is fixed too according to the compositeness 
condition (13) (see detail in \cite {Dong0}). 
Here, we reiterate that since Figs. 2, 3 and 4 are taken into account 
simultaneously, gauge invariance preserves in our effective Lagrangian 
approach. \\

  
\section{Numerical results and conclusions}

{\hskip 0.4cm}
To proceed a numerical calculation, we adopt the parametrization forms of  
the nucleon EM form factors given by Mergell, Meissner and Drechsel 
\cite {MMD}.  Here we follow the numerical technique of \cite {BKT} to 
simplify one of our loop integrations. The loop momentum of the box-type 
Feynman amplitude is parametrized in a such way, that the denominators of 
Green function are $(\mp\kappa +q/2)^2$ for the two photons (see Fig. 5 
where the cross-box diagram is explicitly shown), whereas for the electron (e) 
and the constituent nucleon (N),  they have the forms of 
$(e)=(-\kappa +{\cal K})^2-m_e^2$ and $(p)=(\pm {\cal K}+{\cal P})^2-M_N^2$ 
with 
\begin{eqnarray}
{\cal K}=\frac12(k_1+k'_1)=\frac12K,~~~~~{\cal P}=\frac12(p+p').
\end{eqnarray}
The sign ``-''(``+'') is for the direct (cross-box) diagram in Fig. 5. 
Here, it should be mentioned that the assumption of \cite{BKT} means that 
each of the photons carries approximately half of the transferred 
momentum $q$. It is justified on the bases of Ref. \cite{Gun}. Moreover, 
the assumption also means that a rapid decreasing of the form 
factors is employed such one can neglect the dependence on the loop 
momentum $\kappa$ in the denominators of the photon Green function as well 
as in the arguments of the form factors. This results in ultraviolet 
divergences of the loop momentum integrals with respect to the momentum 
$\kappa$. Thus, a step function $\theta(M_N^2\tau_N-\mid\kappa^2\mid)$ 
is introduced in the loop integration.  It is equivalent to apply a cut-off 
restriction $\mid\kappa^2\mid<M_N^2\tau$. For the photon Green function, 
we have  
\begin{eqnarray}
\frac{1}{\mid \frac q2\pm\kappa\mid^2}<\frac{1}{{\cal P}^2}=\frac{1}
{M_N^2(1+\tau_N)}
\end{eqnarray} 
with $\tau_N=\frac{Q^2}{4M_N^2}$. In our 
calculation for the effect of TPE based on the effective Lagrangian of 
eq. (10), we face two loop-integrations. One is the loop integration with 
respect to the intermediate momentum of $\kappa$, and another is the one with 
respect to the intermediate momentum $k$ of the composites of the deuteron 
(see Fig. 5). To simplify the numerical calculation further, we also use the 
soft approximation for the integral variable $\kappa$ in the first loop 
integration. \\

\begin{figure}[t]
\centering \includegraphics[width=12cm,height=6cm]{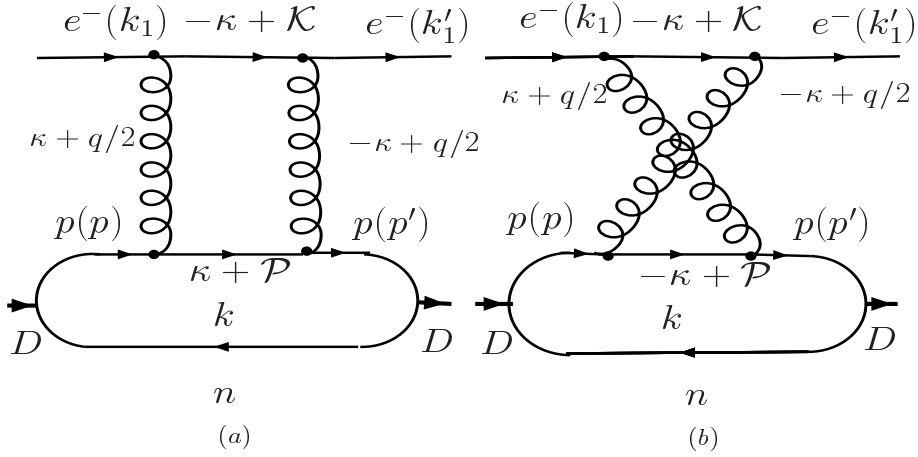}
\caption{\footnotesize Feynman diagrams for two-photon exchange: box 
diagram (a) and crossed box diagram (b).} 
\end{figure}

\par\noindent\par
{\hskip 0.0cm}
Based on the above assumption  and considering the TPE effect shown 
in Figs. 2-4, we may estimate the TPE corrections to the deuteron EM form 
factors in the present effective Lagrangian approach. The effective EM 
interaction Lagrangians have already been given explicitly in 
Refs. \cite {Dong0, 
faesslernew}. With those Lagrangians, we can correctly get the normalization 
conditions for $G_C(0)$ and $G_M(0)$. In Figs. 6-10, we plot our numerical 
results for the contributions of the TPE effect to the deuteron 
electromagnetic form factors of $G_C$, $G_M$, $G_Q$ and to the two additional 
form factors $G_5$ and $G_6$. Two different scattering angles, $\theta$ being 
$\pi/2$ and $\pi/10$, are selected in order to check the $\theta$ dependences 
of the observables. In Figs. 6-8, the ratios stand for 
\begin{eqnarray}
R_{C,M,Q}=\frac{G^{(2)}_{C,M,Q}(s,Q^2)}{G_{C,M,Q}^{exp.}(s,Q^2)},
\end{eqnarray}
where $G^{(2)}_{C,M,Q}(s,Q^2)$ represent for the TPE contributions. The 
individual contributions of Figs. 2, 3, 4 and their sum to the deuteron 
form factors are shown explicitly. $G^{exp.}_{C,M,Q}$ in eq. (18) are 
estimated by the parametrizations of Ref. \cite {GT1} ($\theta$-independent 
form) as the empirical data. The two maximum points in Figs. 6 and 7 are 
due to the two crossing points of the charge $G_C$ and magnetic $G_M$ form 
factors at about $Q^2_{crossing}\sim~0.5~GeV^2$ and 
$Q^2_{crossing}\sim~2.0~GeV^2$. Here, 
different from the form factors of the nucleon, the 
form factors of the deuteron have the crossing points. From Figs. 6-8, one 
cannot explicitly see the $\theta$-dependence of the three ratios, 
since the dependences are strongly suppressed due to the fact 
that the denominators of the ratios in eq. (18) are $\theta$-independent. 
However, the $\theta$-dependences can be seen explicitly in Figs. 9-10 for 
the two new form factors $G_5'=G_5^{(2)}$ and $G_6'=G_6^{(2)}$. It 
should be mentioned that we do not have the extra form factors $G_4'$ 
contributed by the TPE effect as shown in eqs. (6) and (8) since we adopt the 
assumption of \cite {BKT}, where the $\kappa$-dependence terms in the  
numerator are ignored. \\ 
        
In our calculation, we have one parameter $\Lambda_D$ in the correlation 
function. According to the condition that the deuteron is bound as 
$<\mid r^{-2}\mid >\leq 0.02~GeV^2$ \cite {mathiot}, we select a typical value 
for the parameter: $\Lambda_D=0.30~GeV$ which is consistent with the one 
used in Refs. \cite {Dong0, Dong4}. 
Our estimates for the ratios of the deuteron electromagnetic form factors 
of $G_{C,M,Q}'$ tell that the TPE effect is small. To analyze the 
contributions of Figs. 2, 3 and 4, one sees that in the low $Q^2$ region, 
the contributions of Figs. 2 and 3 are smaller than the one of Fig. 4. 
When $Q^2$ increases, the contributions of Figs. 2, 3 and 4 increase too. 
Moreover, the contributions of Figs. 2 and 3 to the form factors of 
$G_{C, Q}$ are always smaller than that of Fig. 4, whereas the one of 
Fig. 2 to $G_M$ becomes compatible to the contribution of Fig.4. Since the 
total contributions of Figs. 2, and 3 to the three conventional EM form 
factors of $G_{C, M, Q}$ are very small, and the $\theta$-dependences of the 
ratios from the TPE corrections are suppressed, it is not easy to directly 
test the TPE effect from the three form factors. \\
 
However, it is expected that one may test the TPE effect from the 
polarizations of deuteron, since the TPE effect is $\theta$-dependent and 
the obtained new form factors $G_{5,6}$ are $\theta$-dependent too. 
Consequently, it is reasonable to find the TPE effect in some polarizations 
and particularly in some angle limit. We know that if only 
one-photon-exchange is considered, the double and single polarization 
observables are 
\begin{eqnarray}
P_{xz}&=&-\tau_D\frac{K_0}{M_D}tan\frac{\theta}{2}G_MG_Q, \nonumber \\
P_z&=&\frac13\frac{K_0}{M_D}\sqrt{\tau_D(\tau_D+1)}tan^2\frac{\theta}{2}
G_M^2.
\end{eqnarray}
Clearly, these two polarizations become vanishing when $\theta$ is very small 
since they are $tan\frac{\theta}{2}$ and $tan^2\frac{\theta}{2}$-dependent, 
respectively. However, when the TPE effect is considered in the small 
angle limit, its contribution is 
\begin{eqnarray}
\delta P_{xz}&\sim &2\tau_D^2cot\frac{\theta}{2}\Bigg 
[2\Big (\frac{G_1}{\tau_D+1}+G_3\Big )Re(G_5')\nonumber \\
&&+\Big(G_1-4G_2+2(\tau_D+1)G_3\Big )Re(G_6')\Bigg ]
\end{eqnarray} 
and
\begin{eqnarray}
\delta P_z&\sim&-\frac{2\tau_D}{3}\sqrt{\frac{\tau_D}{\tau_D+1}}\Bigg [
\Big (3+2(\tau_D+1)tan^2\frac{\theta}{2}\Big )G_2Re(G_5')\nonumber \\
&&+2(\tau_D+1)G_2Re(G'_6)\Bigg ].
\end{eqnarray}
One sees that the TPE corrections to the polarizations do not vanish  
in the limit of $\theta\rightarrow 0$. In Fig. 11, we display the ratios
$R(P_{xz})=\delta P_{xz}/P_{xz}$ for $P_{xz}$, and $R(P_z)=\delta P_z/P_z$ for 
$P_z$ calculated from eqs. (19)-(21). The ratios should behave as 
$1/tan^2(\frac{\theta}{2})$. One sees  that the contributions from Figs. 2 
and 3 are found to be smaller than that of Fig. 4. Moreover, one finds 
the sizeable effect of the two new extra form factors, due to TPE, on the 
polarization observables  $P_{xz}$ and $P_z$. The remarkable 
$\theta$-dependences of the ratios are also displayed in Fig. 11. Therefore, 
a precise measurement of the deuteron polarizations in the small angle limit 
is expected to test the TPE effect. Since the deuteron form factors have 
crossing point $Q^2_{crossing}$, it is also expected to easily find the 
TPE effect at about $Q^2\sim Q^2_{crossing}$. \\
  
\begin{figure}[t]
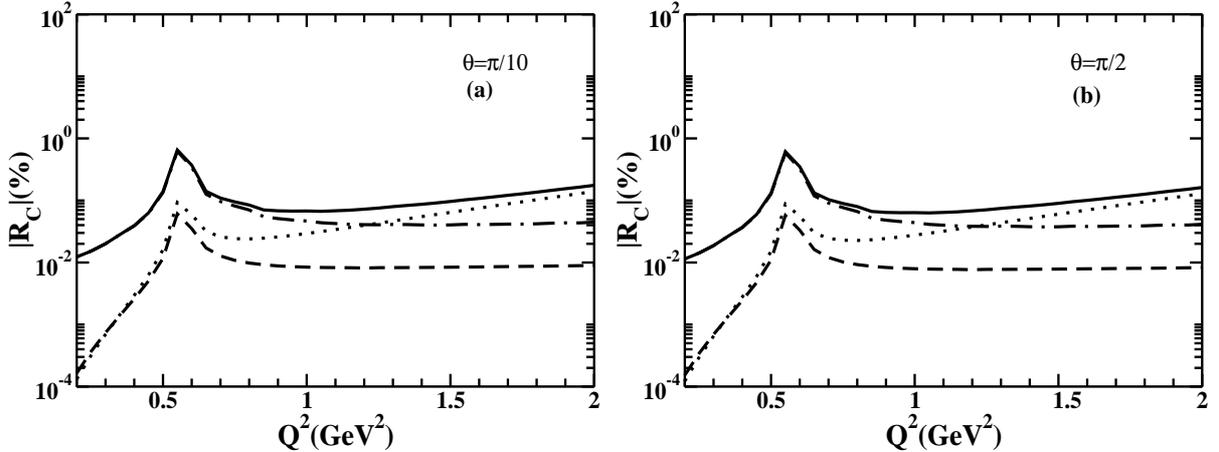

\begin{minipage}[b]{0.5\linewidth}
\centering \includegraphics[width=7.8cm, height=6cm]{rc10v2.eps}
\end{minipage}%
\begin{minipage}[b]{0.5\linewidth}
{\hskip 1.0cm}
\centering \includegraphics[width=7.8cm, height=6cm]{rc02v2.eps}
\end{minipage}
\caption{\footnotesize $\mid R_C\mid(\%)$ for $\theta=\pi/10$ (a) and 
for $\theta=\pi/2$ (b). The dotted, dashed, dotted-dashed and solid curves 
represent the contributions from Figs. 2, 3, 4 and their sum.}
\end{figure}

\begin{figure}[t]
\vspace{2cm}
\begin{minipage}[b]{0.5\linewidth}
\centering \includegraphics[width=7.8cm, height=6cm]{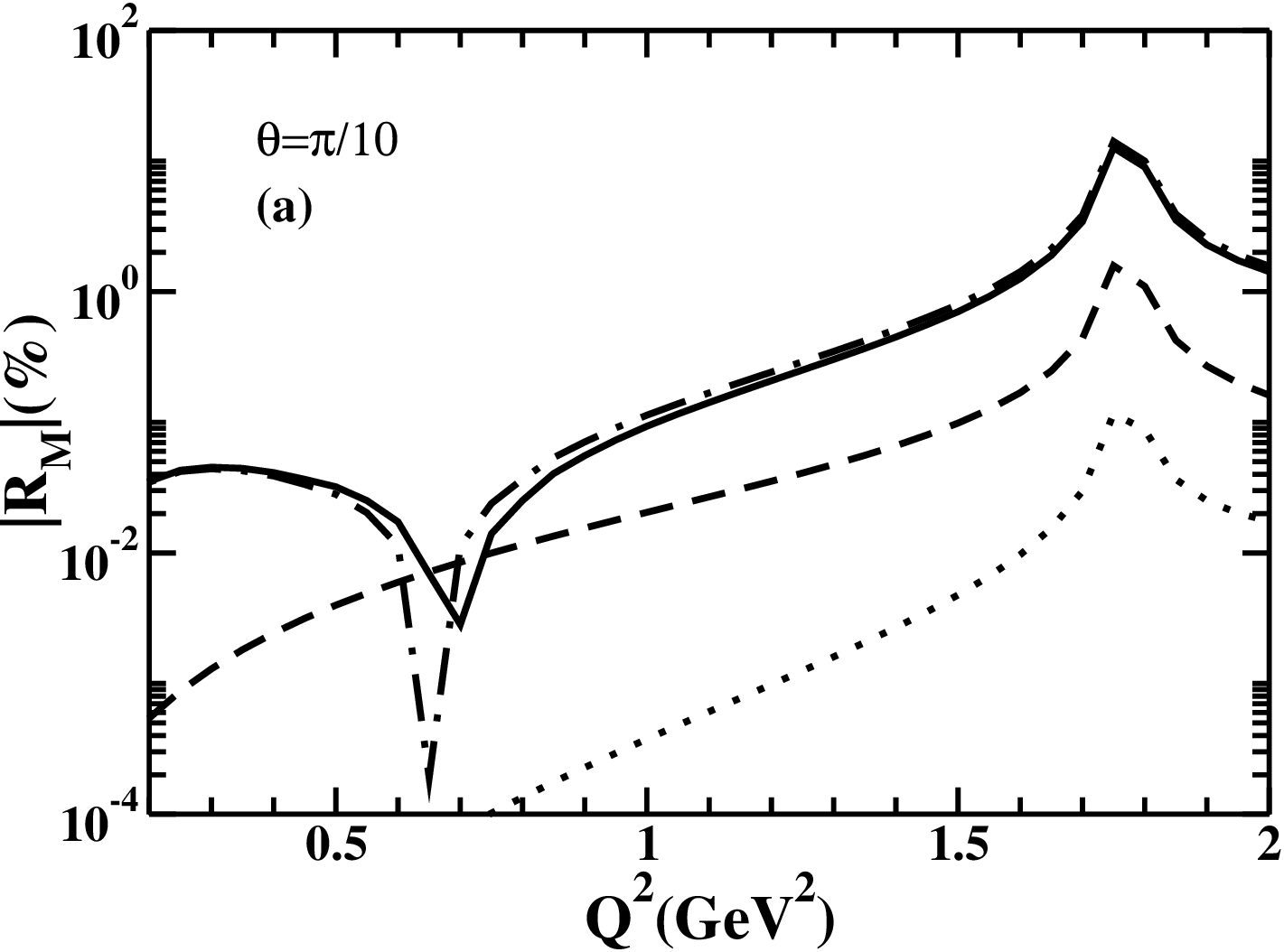}
\end{minipage}%
\begin{minipage}[b]{0.5\linewidth}
{\hskip 1.0cm}
\centering \includegraphics[width=7.8cm, height=6cm]{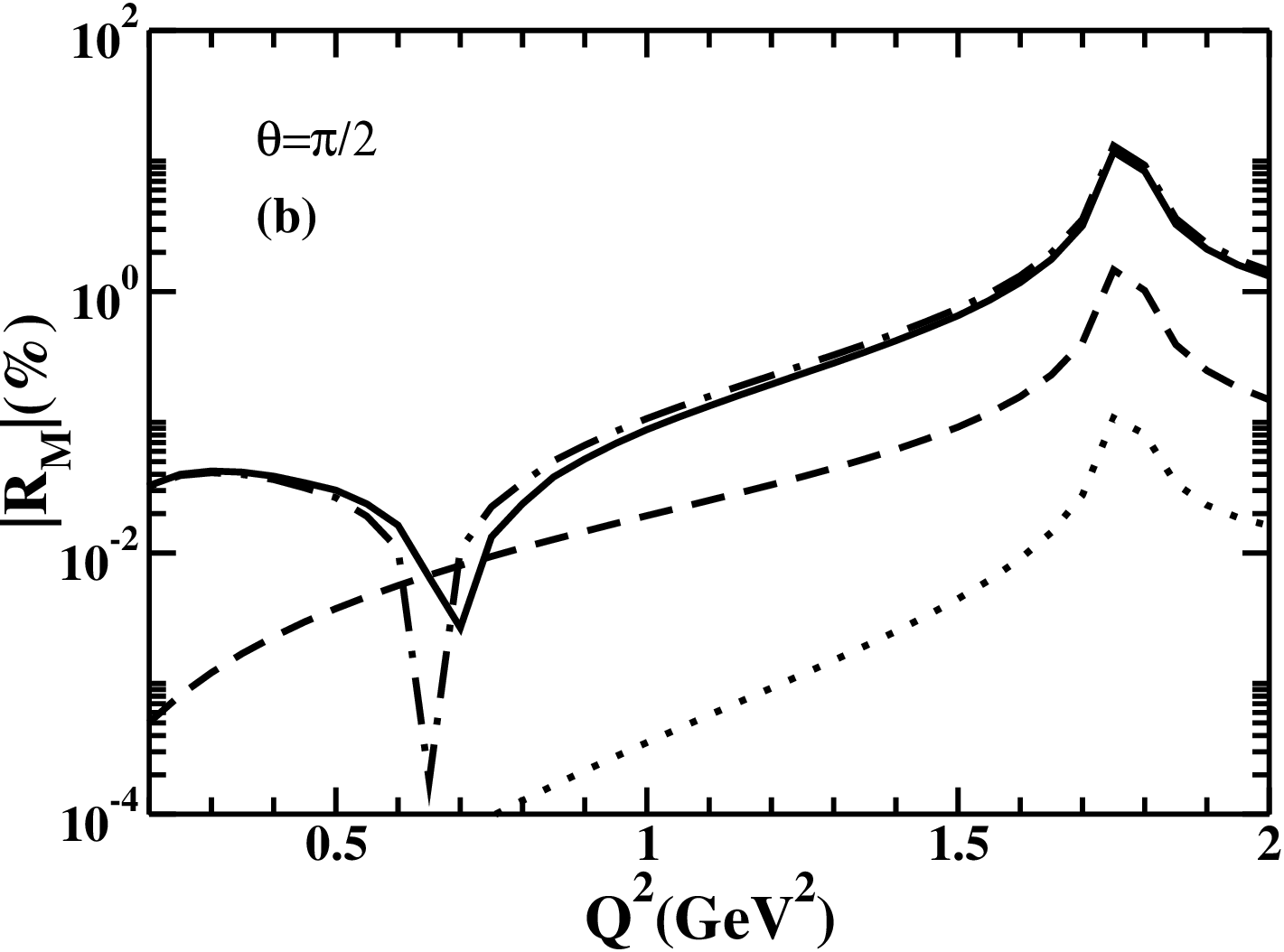}
\end{minipage}
\caption{\footnotesize $\mid R_M\mid(\%)$ for $\theta=\pi/10$ (a) and 
for $\theta=\pi/2$ (b). Notations as Fig. 6}
\end{figure}

\begin{figure}[t]
\vspace{2cm}
\begin{minipage}[b]{0.5\linewidth}
\centering \includegraphics[width=7.8cm, height=6cm]{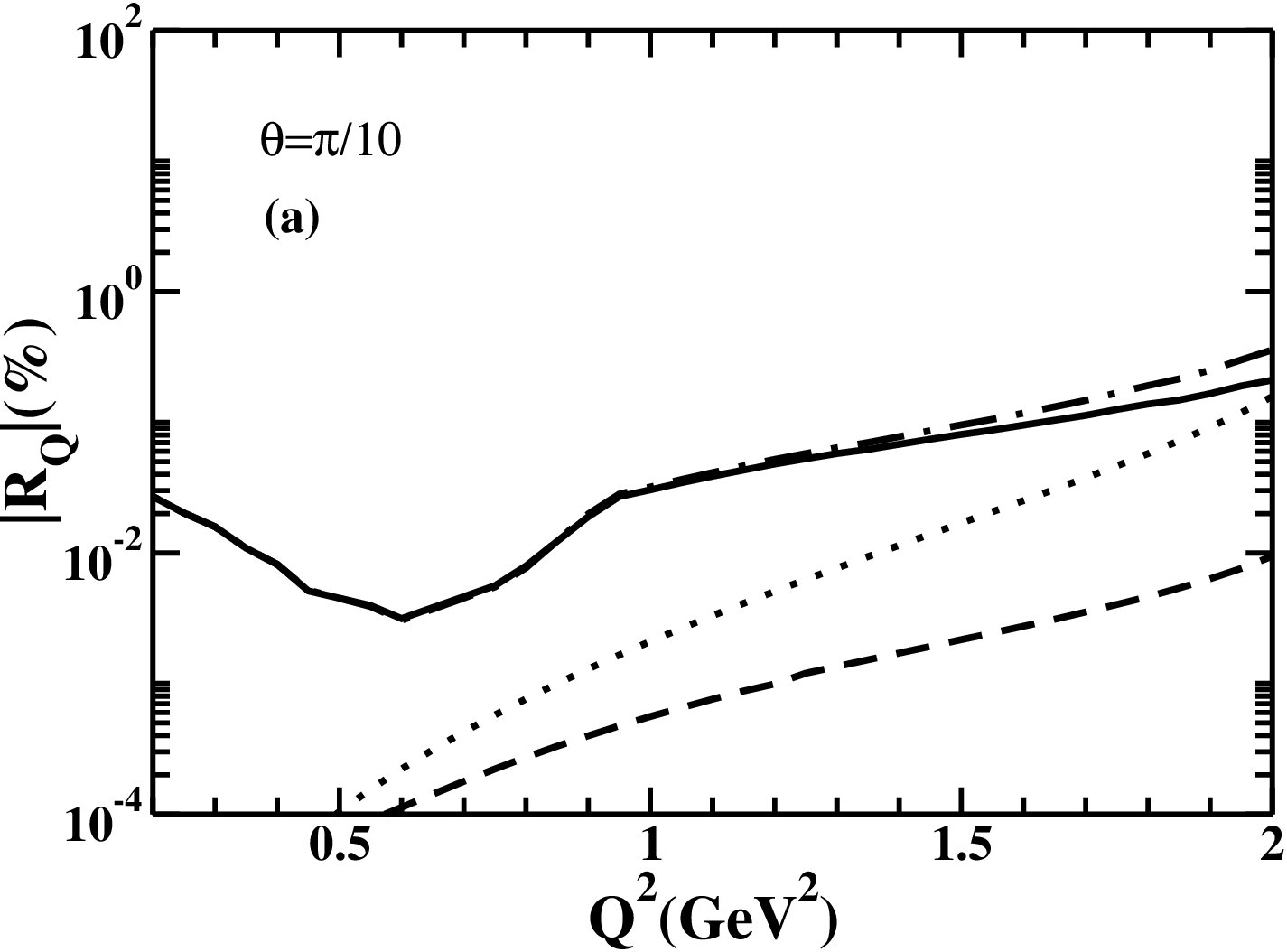}
\end{minipage}%
\begin{minipage}[b]{0.5\linewidth}
{\hskip 1.0cm}
\centering \includegraphics[width=7.8cm, height=6cm]{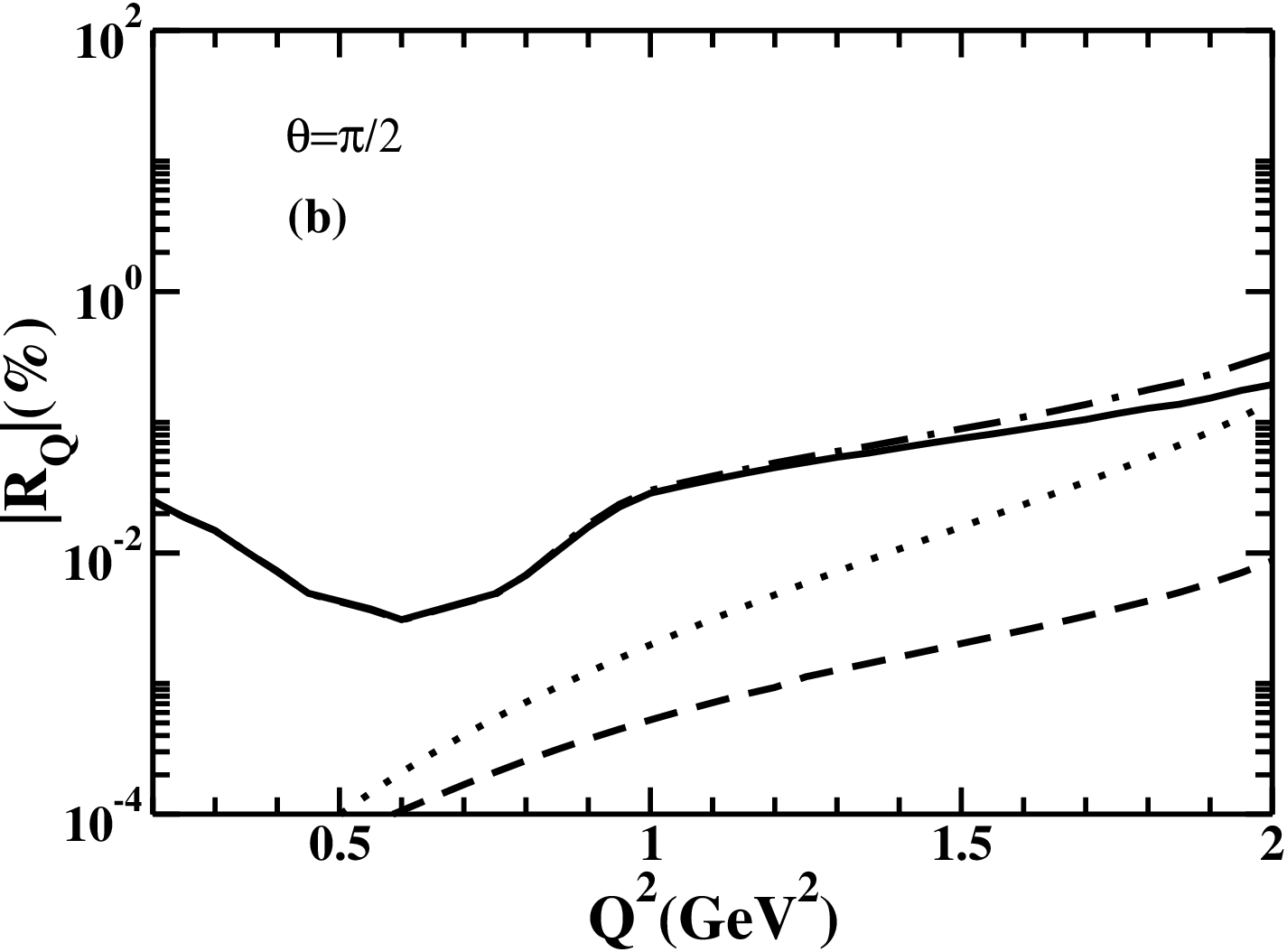}
\end{minipage}
\caption{\footnotesize $\mid R_Q\mid(\%)$ for $\theta=\pi/10$ (a) and 
for $\theta=\pi/2$ (b). Notations as Fig. 6}
\end{figure}

\begin{figure}[t]
\vspace{2cm}
\begin{minipage}[b]{0.5\linewidth}
\centering \includegraphics[width=7.8cm, height=6cm]{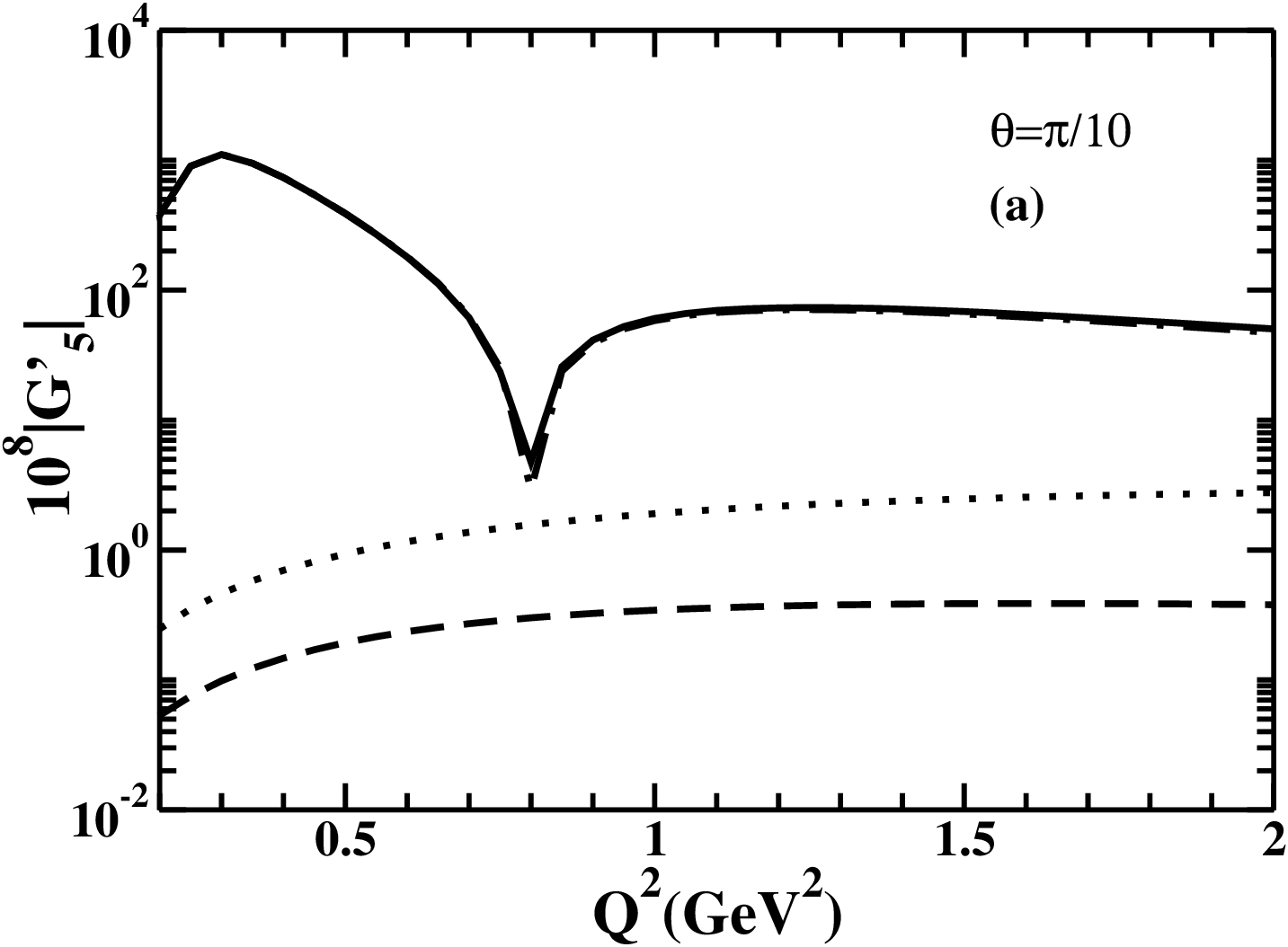}
\end{minipage}%
\begin{minipage}[b]{0.5\linewidth}
{\hskip 1.0cm}
\centering \includegraphics[width=7.8cm, height=6cm]{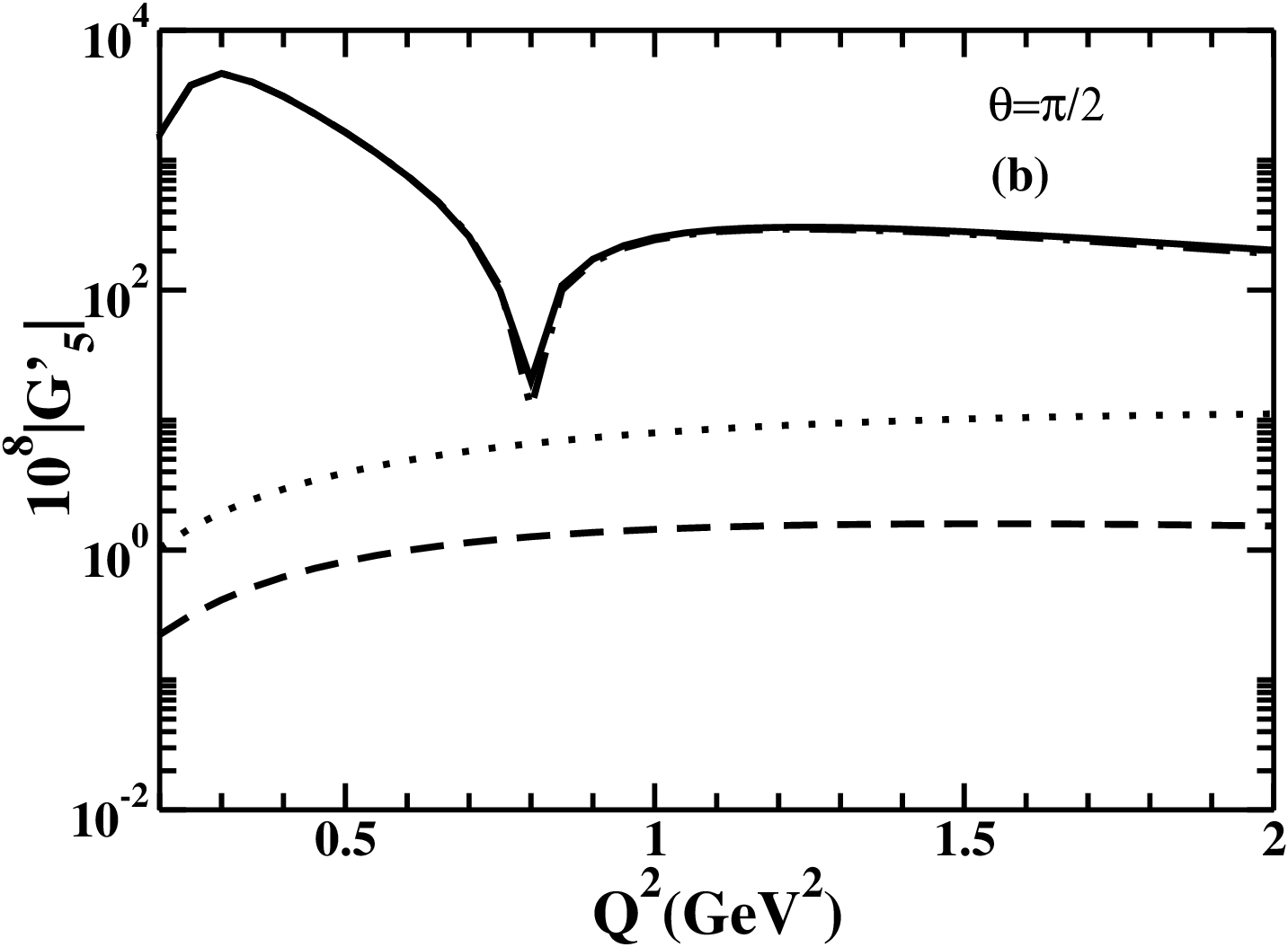}
\end{minipage}
\caption{\footnotesize $10^8\mid G'_5\mid$ for $\theta=\pi/10$ (a) and 
for $\theta=\pi/2$ (b). Notations as Fig. 6}
\end{figure}

\begin{figure}[t]
\vspace{2.0cm}
\begin{minipage}[b]{0.5\linewidth}
\centering \includegraphics[width=7.8cm, height=6cm]{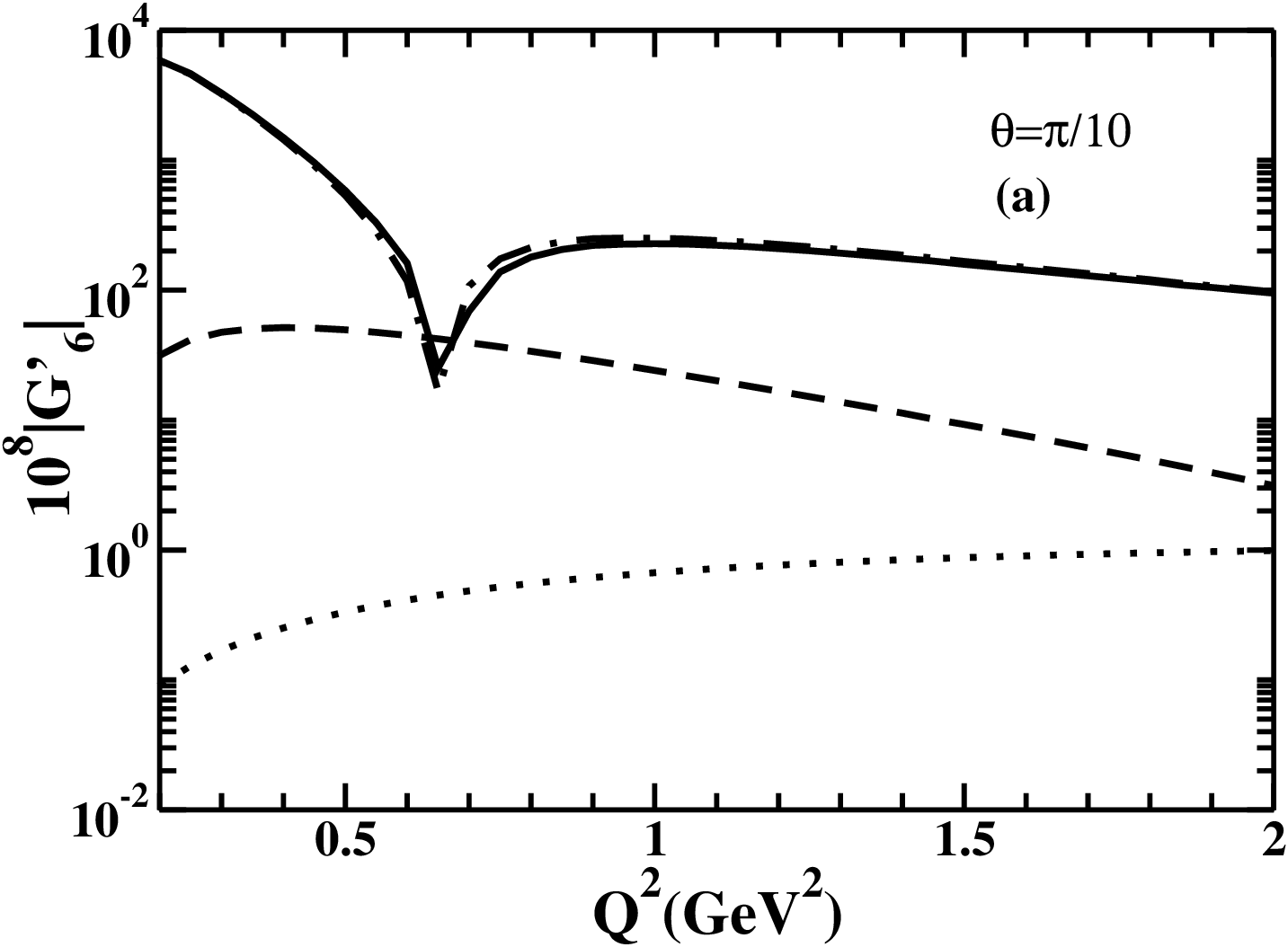}
\end{minipage}%
\begin{minipage}[b]{0.5\linewidth}
{\hskip 1.0cm}
\centering \includegraphics[width=7.8cm, height=6cm]{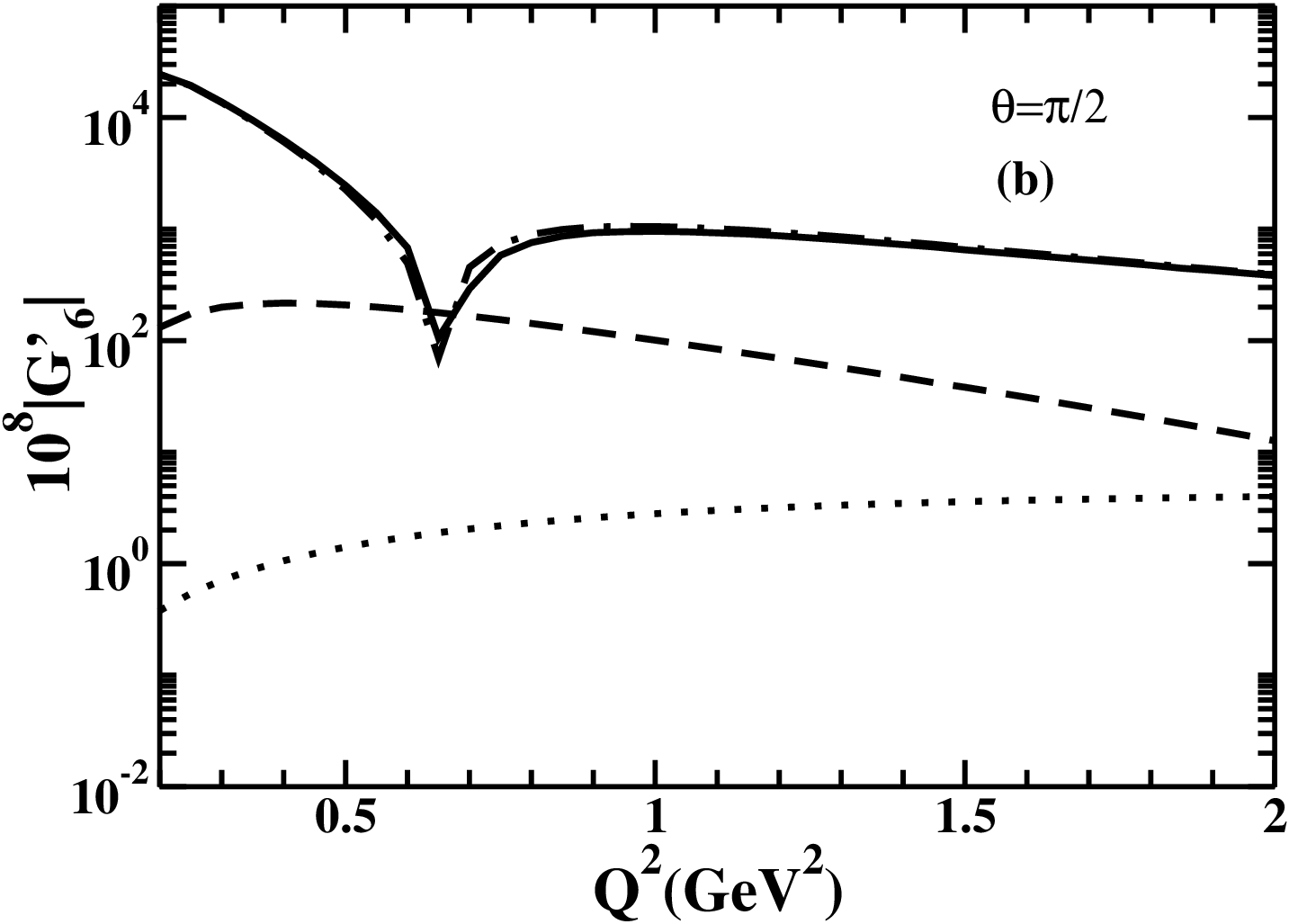}
\end{minipage}
\caption{\footnotesize $10^8\mid G'_6\mid$ for $\theta=\pi/10$ (a) and 
for $\theta=\pi/2$ (b). Notations as Fig. 6}
\end{figure}

Ref. \cite {Dong4} is the first one to numerically estimate part of the TPE 
corrections to the deuteron form factors based on our 
effective Lagrangian approach. In that work, the TPE corrections, 
to the EM form factors of the proton and neutron following the 
formalism of Ref. \cite {Blu0}, are simply employed to study the deuteron 
case. The corresponding TPE effect on the deuteron is shown by Figs. 4(a), 
4(b) and their cross-box diagrams. 
Comparing the present results to those of Ref. \cite {Dong4}, one concludes 
that all the possible TPE corrections are considered in this paper.
Therefore, the present work gives a more systemically and sophisticated 
study of the TPE effect on the deuteron.  Moreover, we directly calculate 
the TPE exchange effect with the assumption of \cite{BKT} in this paper. 
Clearly, the present calculation gives more information about the new 
deuteron form factors since we predict form factors of $G'_{5,6}$ 
simultaneously. The obtained results for the TPE effect are consistent with 
the ones of \cite {Dong4} qualitatively. Finally, one still cannot get any 
information about $G'_4$, this is due to the approximate methods 
we employed here to simplify our numerical loop integration. \\   

{\hskip 0.4cm}To summarize, we are the first to estimate all the TPE 
corrections, as shown in Figs. 2-4, to the conventional form factors of 
the deuteron, $G_{C,M,Q}$ and of $G'_{5,6}$. Our numerical results of the 
TPE contributions tell that $G^{(2)}_{C,M,Q}$ are small (less than $1\%$). 
However, $G_{5,6}'$ are clearly $\theta$-dependent. The two additional form 
factors are expected to be tested in the future measurements of the double 
and single polarization observables of $P_{xz}(T_{21})$ and $P_z$ $(T_{10})$ 
in the small angle limit and at about $Q^2\sim Q^2_{crossing}$. Further work 
for an exactly full calculation of the two-photon exchange effect on the 
deuteron system, without using assumption of \cite {BKT},  
is in progress. \\ 

Finally, this work is also designed to effectively treat direct
electromagnetic interactions to quarks. It should be addressed that 
the present investigation of the two photon exchange mechanism 
recalls a new study of Compton scattering and it is shown that the local 
two-photon coupling to the same quark provides a fixed Regge singularity 
at J=0 \cite {Brodsky}. This subject is beyond the scope of the 
present work. However, it is of a great interest to see the issue for the 
deuteron target in our future work. 

\begin{figure}[t]
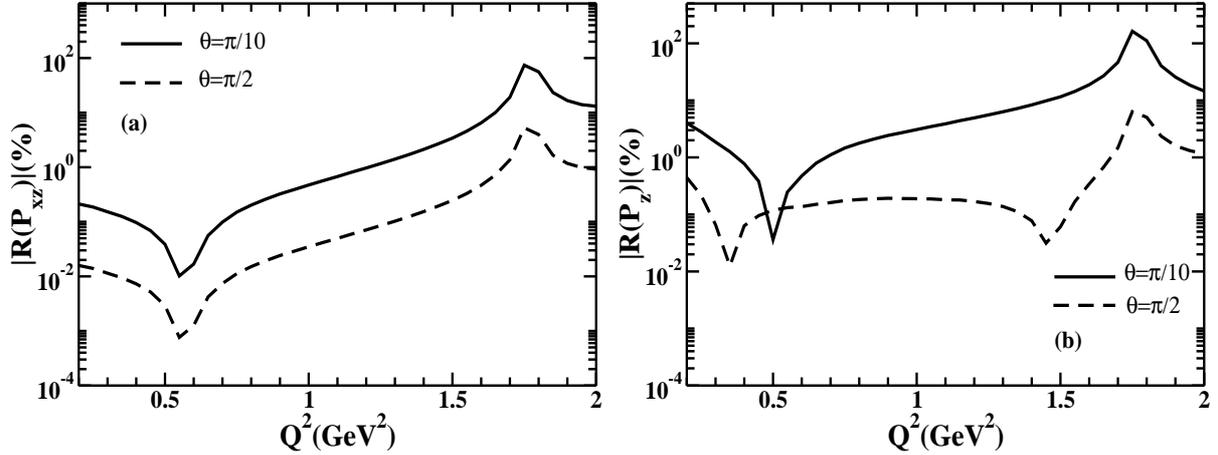

\vspace{2.0cm}
\begin{minipage}[b]{0.5\linewidth}
\centering \includegraphics[width=7.8cm, height=6cm]{rpxzv2.eps}
\end{minipage}%
\begin{minipage}[b]{0.5\linewidth}
{\hskip 1.0cm}
\centering \includegraphics[width=7.8cm, height=6cm]{rpzv2.eps}
\end{minipage}
\caption{\footnotesize Ratios for $P_{xz}$(a) and $P_z$(b).}
\end{figure}

\section{Acknowledgments}
\par\noindent\par\noindent\par
This work is supported  by the National Sciences Foundations grant 
No. 10775148, by CAS grant No. KJCX3-SYW-N2 (YBD), and in part by the 
National Research Council of Thailand through Suranaree University of 
Technology and the Commission of High Education, Thailand.
Discussions with S. N. Yang and Valery E. Lyubovitskij are appreciated.

\end{document}